\documentclass{elsart3}

\def\vec#1{{\bf #1}}

\def\theparagraph    {\arabic{section}.\arabic{paragraph}}
\def\sec#1{\paragraph{#1}}

\usepackage{amssymb}

\begin{document}

\begin{frontmatter}


\title{
 Spontaneous Interlayer Exciton Coherence in Quantum Hall Bilayers at
 $\nu=1$ and $\nu=2$ : A Tutorial\thanksref{label1}}
\thanks[label1]{The author thanks Anna Lopatnikova for teaching
him a great deal about bilayers --- particularly on the topic of
$\nu=2$. He also thanks E. H. Rezayi, M. V. Milovanovic, and G.
Moller for collaborations on the topic of $\nu=1$.  Finally, he
thanks Jim Eisenstein for carefully reading this manuscript.}
\author{Steven H. Simon}
\ead{shsimon@lucent.com}

\title{}


\author{}

\address{}

\begin{abstract}
This tutorial paper reviews some of the  physics of quantum Hall
bilayers with a focus on the case where there is low or zero
tunnelling between the two layers. We describe the interlayer
coherent states at filling factors $\nu=1$ and $\nu=2$ as exciton
condensates and discuss some of the theory associated with these
states.
\end{abstract}

\end{frontmatter}

Over the last decade, a number of extremely important advances ---
both experimental\cite{dassarma,TunExp,Exp2,Kellogg2,Pellegrini}
and
theoretical\cite{dassarma,WenZee,Josephson,Halperin,Bonesteel,OtherApproaches,Yoshioka,Us1,Gunnar,Zheng,brey:99,macdonald:99,burkov:02,Anna}
--- have occurred in the field of quantum Hall bilayer physics. In
this paper, I hope to convey a small subset of the interesting and
important theoretical ideas of this field, deferring many other
fascinating aspects, as well as any detailed discussion the
experiment, to other reviews. The main focus of the paper will be
on the case where the tunnelling between the two layers is
vanishingly small and ``coherence" between the two layers occurs
due to interactions alone.  We will also emphasize how this
physics can be related to the concept of an exciton condensate.

\section{Physics at $\nu=1$}

\sec{The Double Well: } \label{sec:doublewell} Let us start by
imagining a single spinless electron in a double well. We consider
two basis states, $|L\rangle = c^\dagger_L |0 \rangle$ and $|R
\rangle = c^\dagger_R|0 \rangle$ corresponding to the electron in
the left and right wells respectively.  The Hilbert space for a
single electron is just the space of superpositions
$$ u | L\rangle + v |R \rangle = \left( u c^\dagger_L + v
c^\dagger_R \right) | 0 \rangle
$$
 with  the normalization $|u|^2 + |v|^2 =1$. We can also
think of the two-vector $(u, v)$ as an ``isospin" spinor with the
mapping that the $|L\rangle$ state corresponds to isospin-up and
the $|R\rangle$ state corresponds to isospin-down

\unitlength .7mm
\begin{picture}(60,45)(0,40)
\linethickness{0.45mm} \put(10,80){\line(1,0){5}}
\linethickness{0.45mm} \put(15,65){\line(0,1){15}}
\linethickness{0.45mm} \put(15,65){\line(1,0){10}}
\linethickness{0.45mm} \put(25,65){\line(0,1){15}}
\linethickness{0.45mm} \put(25,80){\line(1,0){10}}
\linethickness{0.45mm} \put(35,50){\line(0,1){30}}
\linethickness{0.45mm} \put(35,50){\line(1,0){10}}
\linethickness{0.45mm} \linethickness{0.45mm}
\put(45,50){\line(0,1){30}} \linethickness{0.45mm}
\put(45,80){\line(1,0){5}} \linethickness{0.25mm}
\qbezier(10,65)(12.6,64.99)(14.41,65.59)
\qbezier(14.41,65.59)(16.21,66.2)(17.5,67.5)
\qbezier(17.5,67.5)(18.8,68.81)(20,68.81)
\qbezier(20,68.81)(21.2,68.81)(22.5,67.5)
\qbezier(22.5,67.5)(23.8,66.2)(25,65.59)
\qbezier(25,65.59)(26.2,64.99)(27.5,65)
\qbezier(27.5,65)(28.8,65)(29.41,65)
\qbezier(29.41,65)(30.01,65)(30,65) \linethickness{0.25mm}
\qbezier(30,50)(32.6,49.98)(34.41,51.19)
\qbezier(34.41,51.19)(36.21,52.39)(37.5,55)
\qbezier(37.5,55)(38.8,57.62)(40,57.62)
\qbezier(40,57.62)(41.2,57.62)(42.5,55)
\qbezier(42.5,55)(43.8,52.39)(45,51.19)
\qbezier(45,51.19)(46.2,49.98)(47.5,50)
\qbezier(47.5,50)(48.8,50)(49.41,50)
\qbezier(49.41,50)(50.01,50)(50,50)
\put(20,75){\makebox(0,0)[cc]{$u$}}
\put(41,62){\makebox(0,0)[cc]{$v$}} \linethickness{0.2mm}
\put(55,50){\line(0,1){15}} \put(55,65){\vector(0,1){0.12}}
\put(55,50){\vector(0,-1){0.12}}
\put(62,57.5){\makebox(0,0)[cc]{$2V_\Delta$}}
\put(80,78){\makebox(0,0)[cc]{\small Fig 1: Double Well}}
\put(80,73){\makebox(0,0)[cc]{\small Potential}}

\end{picture}

\noindent A generic single particle Hamiltonian would contain a
tunnelling term and a potential energy difference $V_\Delta$ (or
potential ``bias") between the two wells (as shown in Fig. 1),
giving us a 2 by 2 matrix Schroedinger equation that can be
written as
\begin{equation}
\label{eq:h0}
     ~~~~~~~ \left(%
\begin{array}{cc}
  V_\Delta & -t \\
  -t & -V_\Delta \\
\end{array}
\right) \left(%
\begin{array}{c}
  u \\
  v \\
\end{array}%
\right) = E\left(%
\begin{array}{c}
  u \\
  v \\
\end{array}%
\right)
\end{equation}
where we will assume $t \geq 0$ and real.  For ``balanced" layers
($V_\Delta=0$), the eigenstates are the symmetric state $u =v$
with energy $E_-=-t$ and the antisymmetric state $u=-v$ with
energy $E_+ = t$.  More generally, with $V_\Delta \neq 0$ we will
have \begin{equation}
    ~~~~~~~~E_{\pm}  =\pm \sqrt{V_\Delta^2 + t^2}
\end{equation} with eigenstates such that $|u| \neq |v|$. We will abuse
nomenclature and always refer to the the lower energy state, which
has $v/u$ being positive and real, as being the ``symmetric" state
 and the higher energy
state, which has $v/u$ negative and real, as being the
``antisymmetric" state. In the absence of tunnelling, the energy
of any state is always independent of the complex phase $\phi$ of
$v/u$. However, when tunnelling is added, the energy is lowest
when $v/u$ is positive real, and any attempt to change the complex
phase of $v/u$ will incur an energy cost.

\unitlength .7mm \begin{picture}(105,45)(0,0)
\linethickness{0.3mm} \put(0,20){\line(1,0){60}}
\linethickness{0.3mm}
\multiput(60,20)(0.12,0.12){167}{\line(1,0){0.12}}
\linethickness{0.3mm}  \linethickness{0.3mm}
\multiput(0,20)(0.12,0.12){167}{\line(1,0){0.12}}
\linethickness{0.3mm} \put(20,40){\line(1,0){60}}
\linethickness{0.3mm} \put(5,15){\line(1,0){60}}
\linethickness{0.3mm}
\multiput(65,15)(0.12,0.12){167}{\line(1,0){0.12}}
\linethickness{0.3mm} \linethickness{0.3mm}
\multiput(5,15)(0.12,0.12){42}{\line(1,0){0.12}}
\linethickness{0.3mm} \put(75,35){\line(1,0){10}}
\linethickness{0.3mm}
\multiput(25,35)(1.96,0){26}{\line(1,0){0.98}}
\linethickness{0.3mm}
\multiput(30,20)(1.43,1.43){11}{\multiput(-20,0)(0.12,0.12){6}{\line(1,0){0.12}}}
\linethickness{1.75mm} \put(40,0){\line(0,1){10}}
\linethickness{1.75mm}
\multiput(40,10)(0.12,-0.12){42}{\line(1,0){0.12}}
\linethickness{1.75mm}
\multiput(35,5)(0.12,0.12){42}{\line(1,0){0.12}}
\put(70,5){\makebox(0,0)[cc]{}} \put(70,40){\makebox(0,0)[cc]{}}
\put(50,4){\makebox(0,0)[cc]{\large $B$}}

\put(80,5){\makebox(0,0)[cc]{\begin{minipage}{1in} \small  Fig 2:
A Bilayer in Perpendicular Magnetic
 Field $B$ \end{minipage}}}

\end{picture}

\sec{Introduction to $\nu=1$ Bilayers:}We now turn our attention
to quantum Hall bilayers\footnote{Good general reviews of of many
aspects of $\nu=1$ bilayers are given in Reference
\cite{dassarma}.}. The double well we were considering above now
becomes extended into two parallel 2-dimensional electron sheets,
as shown in Fig. 2.  When a magnetic field $B$ is applied
perpendicular to the layers, the noninteracting electron states
become quantized into Landau levels with energies $(n  + \half)
\hbar \omega_c$ where $\omega_c = e B/m c$ with each Landau level
having $B/\phi_0$ states per unit area  (here $m$ is the electron
mass, $c$ the speed of light, $e$ the electron charge, and $\phi_0
=  2 \pi \hbar e/c$ the flux quantum with $\hbar$ Planck's
constant). At high enough magnetic field, low energy states are
restricted to the Lowest Landau level ($n=0$), and we specify the
2D positional degree of freedom within the lowest Landau level
(LLL) by a single variable $X$. At each possible value of $X$
there are 2 states available for the electron, one in each of the
wells, with tunnelling between the wells analogous to that
discussed above. (For now we neglect the actual spin of the
electron, assuming that the spin is fully polarized, which is a
good assumption in high enough magnetic field). In addition to the
simple physics of an electron in a double well at each position
$X$, now there will also be Coulomb interactions between electrons
at different positions $X$. At filling\footnote{The filling
fraction $\nu$ is defined here to be the {\it total} number of
electrons divided by the number of states per layer, which is
given by $\nu = n \phi_0/B$ with $n$ the density} fraction $\nu=1$
there is precisely one electron in the system for each possible
value of $X$. If the tunnelling between the layers is sufficiently
strong, so that the energy of the symmetric state is much lower
than that of the antisymmetric state, then at each position $X$ we
should fill only the symmetric state and we should leave the
antisymmetric state empty. The multiparticle wavefunction can then
be written as
\begin{equation}
 \label{eq:111}
~~~~~~~~~~~~    \Psi =   \prod_X \left(u c^\dagger_{LX} + v
c^\dagger_{RX}
    \right)|0 \rangle
 \end{equation}
where $(u,v)$ are the coefficients of the symmetric eigenstate of
the double well which must have $v/u$ real and positive.  (As
discussed above, if there is no bias between the two wells, we
would have $u=v$ in the symmetric state). The resulting state (Eq.
\ref{eq:111}) is simply a filled Landau level of electrons in the
symmetric superposition. This wavefunction is  certainly the exact
ground state in the limit where $t$ is large such that the
symmetric state is much lower energy than the antisymmetric state.
Interestingly, as we will discuss further below, it is thought
that this state is also exact in certain circumstances for small,
or even vanishing $t$.

\sec{The $\nu=1$ Quantum Hall Ferromagnet:}

It is extremely interesting to examine the limit where the
tunnelling $t$ between the layers become small.  As discussed
above for the case of the double quantum  well, in this limit the
energy of any superposition state is independent of the complex
phase $\phi$ of $v/u$. Similarly, when $t \rightarrow 0$, the
energy of Eq. \ref{eq:111} becomes independent of this phase.
However, it should not be immediately obvious that Eq.
\ref{eq:111} is the ground state in this limit.  If we recall our
argument that we should fill the symmetric state, but leave the
antisymmetric state empty, we realize that this is no longer a
valid argument in the limit of $t \rightarrow 0$ and
$V_\Delta\rightarrow 0$ where the symmetric and antisymmetric
states both have the same energy. Despite this degeneracy the
electron-electron interaction can strongly favor all of the
electrons being in the {\it same} superposition, in which case Eq.
\ref{eq:111} remains a valid wavefunction, and indeed can be
essentially exact in the limit of small spacing between the two
layers\footnote{\label{foot:1}Changing the layer spacing, of
course, changes the interaction between the layers, and hence can
change the ground state}.

We now think more generally in the case where the interlayer bias
$V_\Delta$ is not zero.   For a single electron in a double
quantum well with $t \rightarrow 0$ and nonzero $V_\Delta$, the
ground state will consist of the electron completely in the well
of lower energy. (What we call the ``symmetric" state --- the
eigenstate with lower energy --- in the limit of $t=0$ is just an
electron in the lower well). However, when one considers a system
of many electrons, due to electron-electron interaction, the
self-consistent ground state will have some density of electrons
in each of the wells.  Indeed, the ground state will again be of
the form of Eq. \ref{eq:111} where the relative magnitude of
$|u|^2$ and $|v|^2$ indicates the relative densities in the left
and right wells respectively. Furthermore, in the limit of $t=0$
the energy of Eq. \ref{eq:111} must remain
independent\footnote{When $t=0$, the interacting Hamiltonian has a
precise gauge symmetry which can be expressed as $c^\dagger_{RX}
\rightarrow e^{i \phi} c^\dagger_{RX}$ with the same $\phi$ at all
points in space.  This guarantees the independence of the energy
of the wavefunction Eq. \ref{eq:111} as the phase $\phi$ is varied
changed.} of the phase $\phi$ of $v/u$ so long as this phase is
chosen the same at all points $X$ in space.

As mentioned above, the two vector $(u,v)$ with $|u|^2 + |v|^2 =1$
can be thought of a spin-${1}\over{2}$ iso-spinor. In the language
of spin\footnote{The expectation of the components of the spin are
given by $\langle S_i \rangle= (u^*, v^*) \sigma_i (u, v)^T$ with
$\sigma_i$ the Pauli spin matrices and $i=x,y,z$.}, the
$z$-component $\langle S_z \rangle \propto |u|^2 - |v|^2$
represents the density imbalance between the layers. This quantity
is fixed by the interlayer bias $V_\Delta$ and the interaction
between electrons. Variation of $\langle S_z \rangle$ from its
preferred value will cost ``capacitive" energy. On the other hand,
the azimuthal angle of the spin is given by the phase $\phi$ of
$v/u$. As discussed above, when $t=0$, no particular value of this
phase is preferred, so long as the phase (or equivalently the
iso-spin direction) is chosen the same at all points in space.
This alignment of the iso-spin vector at all points in space gives
the wavefunction Eq. \ref{eq:111} the name ``quantum Hall
ferromagnet."      When the system chooses a particular value of
$\phi$ among all of the physically equivalent possibilities we say
there has been a spontaneous breaking of symmetry.

In this limit of no tunnelling between the two layers, there is a
thus a one parameter family of physically equivalent ground states
paramterized by the phase $\phi$.  This phase is a quantum
mechanical variable that is conjugate to the difference in the
number of particles between the two layers.  In Eq. \ref{eq:111}
the phase $\phi$ of $v/u$ is well defined, but the number of
particles in each layer is maximally uncertain\footnote{It is an
easy exercise show that for the wavefunction of Eq. \ref{eq:111}
the variance in $N_L- N_R$ is $2 |u v| N^{1/2}$.}. Conversely, by
making the phase completely uncertain, we can construct a
wavefunction that precisely specifies the number of particles in
each layer. To do so, we integrate over the phase freedom
\begin{eqnarray} \label{eq:Nfixed}
    \Psi_{N_R} &=& \int \frac{d\phi}{2 \pi} \, e^{-i N_R \phi} \prod_X \left(u c^\dagger_{LX} +
    v e^{i \phi}
c^\dagger_{RX}
    \right)|0 \rangle
\end{eqnarray}
and obtain a wavefunction with precisely $N_R$ electrons on the
right side and $N_L = N - N_R$ on the left  with $N$ the total
number of electrons, which at $\nu=1$ is equal to the number of
spatial orbitals.

As in all cases, when a spontaneously broken symmetry  (here
$\phi$) can vary as a function of position, there must be a
Goldstone mode corresponding to long wavelength, low energy
variations of the broken variable --- analogous to the spin wave
of a ferromagnet. We can study this possibility by proposing a
wavefunction that allows the phase to vary locally
\begin{equation}
\label{eq:111X}
  ~~~~~~  \Psi(t)  =  \prod_X \left(u c^\dagger_{LX} + v e^{i \phi(X,t)}
c^\dagger_{RX}
    \right)|0 \rangle
\end{equation}
and without loss of generality, we can now take $u$ and $v$ both
real and positive.  As we will argue below, $\phi$ is a superfluid
phase for this system, where the superfluid mode corresponds to
equal and opposite currents propagating in the two opposite
layers.

The local order parameter for this symmetry broken state is
\begin{equation} \label{eq:order}
~~~~~~   \psi_X = \langle c^\dagger_{LX}
c^{\phantom{\dagger}}_{RX} \rangle = u v \, e^{i \phi(X)}
\end{equation}
which is the expectation of an operator that takes a particle out
of the right layer and puts it in the left layer.  It is quite
interesting that the expectation of this operator can be large
(order unity) even for infinitesimally small tunnelling between
the two layers so long as the ground state is still described by
Eq. \ref{eq:111}.

Further, we can write an expression for the tunnelling current
operator between the two layers as a function of position
\begin{equation}
\label{eq:intercurr}
  ~~~~~~ j_{LRX} =      -i t (c^\dagger_{RX} c^{\phantom{\dagger}}_{LX} - c^\dagger_{LX}
  c^{\phantom{\dagger}}_{RX}) \end{equation}
  whose expectation is given by \begin{equation}
  ~~ \left\langle j_{LRX} \right\rangle = -i t (\psi^*_X - \psi_X) = 2 t u v
  \sin[\phi(X)]
\end{equation}
which appears to be similar to a Josephson current --- depending
on the superfluid phase, and not in any explicit way on the
potential bias between the two layers.  Indeed, in experiments a
greatly enhanced tunnelling current is observed at very low
bias\cite{TunExp}. However, this enhanced current also has some
notable differences with classic Josephson
tunneling\cite{Josephson} and there appears to be no tunnelling
current at precisely zero bias.

We note that when the number of particles in each layer is
strictly conserved (when $t$ is zero, and the system is isolated
from the leads), as in Eq. \ref{eq:Nfixed}, the expectation in Eq.
\ref{eq:order} is clearly zero.  Nonetheless, essentially the same
broken symmetry still exists, which can be seen by examining a
slightly more complicated quantity such as
\begin{equation} ~~~~ G_{X,X'} = \langle c^\dagger_{LX}
c^{\phantom{\dagger}}_{RX} c^\dagger_{RX'}
c^{\phantom{\dagger}}_{LX'}\rangle  \end{equation} which conserves
the particle number in each layer, is equal to
$\psi_X^{\phantom{*}} \psi^*_{X'}$ for the wavefunction Eq.
\ref{eq:111X}, and is nonzero (order unity) even for the
wavefunction Eq. \ref{eq:Nfixed}.

\sec{Exciton Language:} \label{sec:exciton}

The  state Eq. \ref{eq:111} can equally well be thought of as a
condensate of excitons\cite{WenZee}.  To see this, we imagine an
effective vacuum state given by one of the quantum wells being
completely empty and the other being a completely full LLL. For
example, if we write \begin{equation}~~~~~~~~~ |\mbox{Filled-Left}
\rangle = \prod_X c^\dagger_{LX} |0\rangle \end{equation}  we then
have the wavefunction of Eq. \ref{eq:111} being given by
\begin{equation}
 \label{eq:BCS} ~~~~~~ \Psi = \prod_{X} \left( u + v c^{\dagger}_{RX}
c^{\phantom{\dagger}}_{LX}
    \right) |\mbox{Filled-Left} \rangle
\end{equation}
This then appears essentially identical to an exciton
condensate\cite{ExcitonCondensate} in the standard BCS form where
the excitons $c^\dagger_{R} c^{\phantom{\dagger}}_L$ are now made
by taking an electron out the filled left band and placing it in
the previously empty right band.   The excitons should then be
thought of as an electron in the right layer bound to a hole in
the left layer.   Superfluid motion of these bound objects now
clearly corresponds to counter-propagating charge currents as
claimed above. Indeed, experiments have observed vanishing
resistance for counter-propagating currents\cite{Kellogg2}.

It is interesting to note that, as usual with wavefunctions of BCS
form, one can rewrite Eq. \ref{eq:BCS} in a form reminiscent of a
coherent state of bosons\footnote{In going from Eq. \ref{eq:BCS}
to Eq. \ref{eq:coherent} we use the fact that $b^\dagger_X
b^\dagger_X  = 0$ due to the Fermi statistics of the underlying
electrons.  Thus we should think of the excitons as being
hard-core bosons.}
\begin{equation} \label{eq:coherent}
  ~~~~~~~~~~    \Psi = \prod_{X} e^{(v/u) b^\dagger_X} |\mbox{Filled-Left} \rangle
\end{equation}
where $b^\dagger_X = c^\dagger_{RX} c^{\phantom{\dagger}}_{LX}$ is
the exciton-boson.  Thus, we speak of this wavefunction as being a
``coherent" state.    The term ``spontaneous coherence" is often
used for the case of zero tunnelling to mean that the global
formation of this coherent state is purely due to
electron-electron interaction. This is in contrast with the case
of finite tunnelling where the symmetric superposition (and hence
wavefunctions of the form of Eq. \ref{eq:111}) is energetically
favored even in the absence of interaction.

We note that, similarly to the above manipulations, one could have
started with a filled right LLL band and created excitons of the
form $c^\dagger_{L} c^{\phantom{\dagger}}_R$ to obtain the very
same state as follows
\begin{equation}
 \label{eq:BCS2} ~~~~~~ \Psi = \prod_{X} \left( u c^{\dagger}_{LX}
c^{\phantom{\dagger}}_{RX} + v
    \right) |\mbox{Filled-Right} \rangle
\end{equation}
In this language, the exciton-boson is instead written as $\tilde
b^\dagger_X = c^{\dagger}_{LX} c^{\phantom{\dagger}}_{RX}$ which
is just $b_X$ in the above definition we used in describing Eq.
\ref{eq:BCS}. Thus, creating an exciton in the language of Eq.
\ref{eq:BCS2} is the same as annihilating an exciton in the
language of Eq. \ref{eq:BCS}. This duality should not surprise us
given the particle-hole symmetry of a half-filled Lowest landau
level. Here, we have precisely half of the available states filled
and it is equally valid to think about these states as being
half-full of electrons or half-full of holes.

\sec{First Quantized Language:}  \label{sec:first} Bilayer
wavefunctions were first proposed in the language of first
quantized trial wavefunctions\cite{Halperin}.  The wavefunction of
Eq. \ref{eq:111} is described as a filled Landau level of
electrons in a particular superposition described by the spinor
$(u,v)$.  We can write this wavefunction in first quantized form
as
\begin{eqnarray} \nonumber ~~~~~~ ~~~ \Psi &=& \prod_i \left( u |L_i
\rangle + v |R_i \rangle \right) \otimes  \\ & & ~~~~~ ~~~~~~~
\Psi_{FilledLLL}(\vec r_1, \ldots, \vec r_N)
\end{eqnarray}
where the second term is the wavefunction for the positional
degree of freedom and the first term is the wavefunction of the
iso-spin or layer degree of freedom that puts each electron in the
$(u,v)$ superposition.

Of course, when the tunnelling between layers is strictly zero, as
discussed above, the number of electrons in each layer should be a
conserved quantity.  As discussed above in Eq. \ref{eq:Nfixed}
this restriction can be enforced properly by integrating over the
the phase degeneracy $\phi$ of $v/u$ to obtain
\begin{equation}
~~~~~~~~~ \Psi_{N_R} = \hat{\mathcal P}_{N_R}
\Psi_{FilledLLL}(\vec r_1, \ldots, \vec r_N)
\end{equation}
where $\hat {\mathcal P}_{N_R}$ is a projection operator that
enforces that precisely $N_R$ electrons should be on the right
side.

We now turn to writing the wavefunction out explicitly.  In radial
gauge, the single electron spatial eigenstates of the lowest
Landau level are given by $\varphi_m(\vec r) = z^{(m-1)}
e^{-|z|^2/(4 \ell^2)}$ where $\ell$ is the magnetic length, $m$
ranges from 1 to the number of orbitals and $z = x + i y$ is the
position $\vec r = (x,y)$ written as a complex number.  The
wavefunction of a filled Lowest Landau level is then given by the
Slater, or Vandermonde, determinant
\begin{eqnarray}
~~~~    \Psi_{FilledLLL} &=& \det[\varphi_m(\vec r_n)] = \det[z_n^{(m-1)}] \\
&=& \prod_{i <j} (z_i - z_j) \label{eq:slat}
\end{eqnarray}
where we have dropped the Gaussian factors for simplicity of
notation.

In the limit where there is strictly no tunnelling between the
layers, we need only project this wavefunction such that there are
precisely $N_R$ particles on the right.   In this limit of no
tunnelling, the particles on the right and the left become
distinguishable particles, in which case it is conventional to
write the coordinates of the $N_R$ particle on the right as $z_i$
and the coordinates of the $N_L=N-N_R$ particles on the left as
$w_i$. Performing the projection, we then obtain an explicit
expression for wavefunction Eq. \ref{eq:Nfixed}
\begin{eqnarray} \nonumber ~~~~~~~~\Psi_{N_R} &=&  \prod_{i <j \leq N_R}
 (z_i - z_j)  \prod_{ i <j \leq N_L}
(w_i - w_j)  \\ & &  ~~~~~~ \prod_{i \leq N_R ; j \leq N_L} (z_i -
w_j) \label{eq:111realspace}
\end{eqnarray}
which is the form first written down over twenty years
ago\cite{Halperin}.   Interesting generalizations of this
wavefunction that may occur at other filling fractions fields
include the possibility of raising the first two factors to some
odd power $m$ and the last factor to some power $n$ to generate
the so-called $mmn$-state. For this reason, the wavefunction Eq.
\ref{eq:111realspace}, and sometimes the wavefunctions of the form
Eq. \ref{eq:111}, are often called the 111-state.

  Naturally, as required by Fermi statistics,
  due to the two factors on the first line of Eq. \ref{eq:111realspace}, this
wavefunction vanishes when any two particles in the same layer
come together (any two $z$'s  come to the same position or any two
$w$'s come to the same position). The interesting physics of this
wavefunction is from the factor on the second line
--- the term that makes the wavefunction also vanish when any $z$
comes to the same location as any $w$. This means that wherever
there is a particle in one layer, the space directly opposite it
in the other layer cannot have an electron, or equivalently must
have a hole. This binding of electron in one layer to a hole in
the other is, of course, the same exciton that was discussed above
in Section \ref{sec:exciton} in second quantized language.

\sec{Quantum Disordering the 111 State, In Brief:}

Although the above described 111 state (Eq. \ref{eq:111} or Eq.
\ref{eq:111realspace}) is clearly the exact ground state at high
enough tunnelling strength, at small tunnelling strength it is not
so clear that this should be exact.   As mentioned above, it turns
out that for Coulomb interactions, even at $t=0$ the 111 state is
indeed an exact ground state when the distance between the two
layers goes to zero $\!\!\!{}^{\mbox{\tiny \ref{foot:1}}}$. In the
opposite limit, however, where the layers are very very far apart,
the interaction between the layers drops to zero and we should
clearly have two separate and uncorrelated layers. In the
``balanced" case, where the density in the two layers is the same,
we expect that we should end up with two independent layers (not
interacting with each other) each with half the density
($\nu=1/2$). Fortunately, the ground state of such a $\nu=1/2$
single layer system is well understood to be described as a
composite fermion fermi liquid\footnote{It has been
proposed\cite{Bonesteel} that for arbitrarily small (but nonzero)
interaction strength between the two layers, there could still be
a Cooper pairing instability of the composite fermion fermi seas,
although it would certainly occur at exponentially low
temperature.}\cite{Composite}
--- a state with very strong electron-electron correlations within
the single layer.  Indeed, the strong electron-electron
correlation within the layer can be thought of as binding of
electron to a correlation hole within the same
layer\cite{Composite,Read}. Thus, the crossover between the 111
state at small layer spacing and the composite fermion liquid at
large layer spacing is one where the inter-layer excitons are
replaced by intra-layer electron-hole binding. Simultaneously, we
expect that, as the layer spacing is increased, the coherence
order parameter $\psi$ (Eq. \ref{eq:order}) must fall from its
finite value in the 111 phase to a zero value in the composite
fermion liquid phase. Exact diagonalization studies
\cite{Yoshioka} indeed show the continuous reduction of the order
parameter as the layer spacing is increased.   This reduction of
the order in the ground state as the parameter of layer spacing is
changed, is sometimes known as ``quantum disorder."

It is very clear experimentally, in both tunnelling and drag
experiments, that some sort of crossover or phase transition
between two very different limiting states is
occurring\cite{dassarma,TunExp,Exp2,Kellogg2}. However, the nature
of this crossover is still a topic of theoretical
discussion\cite{dassarma,OtherApproaches,Yoshioka,Us1,Gunnar}. One
proposal (made by the current author and
collaborators\cite{Us1,Gunnar}) is to describe the transition by a
set of trial wavefunctions that allow a variable number of inter-
versus intra-layer excitons. The crossover is then easily
described by varying the amount of inter- versus intra layer
binding.  Overlaps of these trial states with exact ground states
is found to be quite good for small systems\cite{Us1}.

\section{Some Physics at $\nu=2$}
\setcounter{paragraph}{0}

\sec{The Double Well With Spin:}

We now consider the more
rich\cite{Zheng,brey:99,macdonald:99,burkov:02,Anna} physics of
bilayers at $\nu=2$. Here, we will consider both the layer degree
of freedom (which we have called the iso-spin) as well as the
actual spin degree of freedom of the electron. Resorting to our
simple model of a single electron problem as above in section
\ref{sec:doublewell}, we now study a single electron {\it with
spin} in a double well. As discussed above, the spatial
wavefunction has two eigenstates: the lower energy symmetric $|S
\rangle$ state and the higher energy antisymmetric $|A\rangle$
state with energies $\pm \sqrt{t^2 + V_\Delta^2}$. When the
electron has spin, the spin can be in a spin-up eigenstate, which
we write as $|\uparrow \rangle$ or a spin-down state, which we
write as $| \downarrow \rangle$. In the presence of a magnetic
field, the spin couples to the magnetic field and there will be a
Zeeman energy splitting between the up and down states which is
usually written as $2 V_z=g \mu B$. Thus, the spinful electron in
the double well has a total of four possible eigenstates (See Fig.
3) given by $|S\uparrow~\!\!\!\rangle$,
$|S\downarrow~\!\!\!\rangle$, $|A\uparrow~\!\!\!\rangle$ and $|A
\downarrow~\!\!\!\rangle$  with energies  \begin{equation}
~~~~~~~~E = \pm \sqrt{t^2 + V_\Delta^2} \pm V_z\end{equation}
where the first $\pm$ is $-$ for the symmetric state and $+$ for
the antisymmetric state, and the second $\pm$ is $-$ for the
spin-down state and $+$ for the spin-up state.   Thus, the
symmetric, spin-down state $|S \downarrow\rangle$ is always the
lowest energy state, and is always the ground state of the single
spinful electron in the double quantum well.

\vspace*{.2in}
\unitlength .6mm
\begin{picture}(90,90)(0,0)
\linethickness{0.4mm}
\multiput(30,65)(0.24,0.12){167}{\line(1,0){0.24}}
\linethickness{0.4mm}
\multiput(30,60)(0.24,-0.12){83}{\line(1,0){0.24}}
\linethickness{0.4mm}
\multiput(50,50)(0.24,0.12){83}{\line(1,0){0.24}}
\linethickness{1mm}
\multiput(30,40)(0.24,0.12){83}{\line(1,0){0.24}}
\linethickness{1mm}
\multiput(50,50)(0.24,-0.12){83}{\line(1,0){0.24}}
\linethickness{1mm}
\multiput(30,35)(0.24,-0.12){167}{\line(1,0){0.24}}
\linethickness{0.3mm}
\multiput(10,0)(0,1.98){46}{\line(0,1){0.99}}
\put(10,92){\vector(0,1){0.12}} \linethickness{0.3mm}
\multiput(10,0)(1.97,0){45}{\line(1,0){0.99}}
\put(102,0){\vector(1,0){0.12}} \linethickness{0.1mm}
\multiput(50,0)(0,1){49}{\line(1,0){0.24}}
\multiput(15,50)(1,0){60}{\line(0,1){0.24}}

\put(0,50){\makebox(0,0)[cc]{$E_F$}}

\put(22,66){\makebox(0,0)[cc]{$|A\!\uparrow\rangle$}}

\put(22,59){\makebox(0,0)[cc]{$|A\!\downarrow\rangle$}}

\put(22,41){\makebox(0,0)[cc]{$|S\!\uparrow\rangle$}}

\put(22,34){\makebox(0,0)[cc]{$|S\!\downarrow\rangle$}}

\put(80,85){\makebox(0,0)[cc]{$|A\!\uparrow\rangle$}}

\put(80,60){\makebox(0,0)[cc]{$|S\!\uparrow\rangle$}}

\put(80,40){\makebox(0,0)[cc]{$|A\!\downarrow\rangle$}}

\put(80,15){\makebox(0,0)[cc]{$|S\!\downarrow\rangle$}}

\linethickness{0.1mm} \put(25,10){\line(1,0){20}}
\put(25,10){\vector(-1,0){0.12}} \linethickness{0.1mm}
\put(55,10){\line(1,0){20}} \put(75,10){\vector(1,0){0.12}}
\put(35,5){\makebox(0,0)[cc]{{\bf Singlet}}}

\put(65,5){\makebox(0,0)[cc]{{\bf Ferro}}}

\put(0,85){\makebox(0,0)[cc]{Energy}}

\put(75,-7){\makebox(0,0)[cc]{$V_z / \sqrt{t^2 + V_\Delta^2}$}}

\put(55,-25){\makebox(0,0)[cc]{\begin{minipage}{2.7in} \small Fig
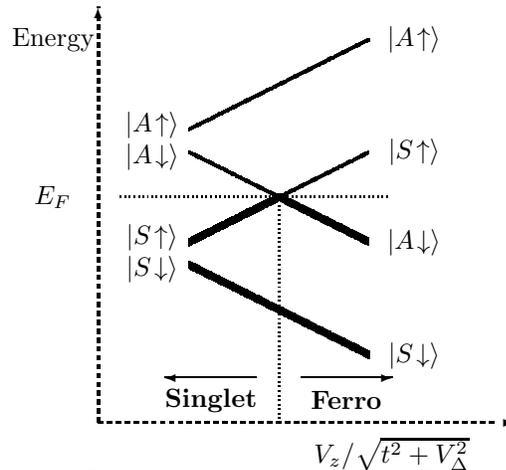
3: Schematic energy diagram for 2 electrons in a double quantum
well as a function of the ratio of the Zeeman energy $V_z =g \mu
B/2$ to the symmetric-antisymmetric splitting $\sqrt{t^2 +
V_\Delta^2}$. The thick lines are filled states below the Fermi
energy
\end{minipage}}}

\end{picture}

\vspace*{1.1in}

We now turn our attention to the more interesting case where there
are two spinful electrons in the double quantum well, and for
simplicity, let us turn off the interaction between the two
electrons.  Now we should fill the two lowest of the four single
electron states in the well.  Again, the symmetric  spin-down
state $|S \downarrow\rangle$ is always the lowest energy state,
and should always be filled.   If $\sqrt{t^2 + V_\Delta^2} > V_z$,
the second lowest state is the symmetric, spin-up state. Filling
the lowest two states in this case (spin-up and spin-down
symmetric states) we obtain the so-called the ${\rm \bf S}$ state,
where $\rm \bf S$ stands for ``Symmetric" or ``Singlet". On the
other hand, if $\sqrt{t^2 + V_\Delta^2} < V_z$, the second lowest
state is the antisymmetric, spin-down state. Filling the lowest
two states in this case (symmetric and antisymmetric spin-down) we
obtain the $\rm \bf F$ or ferromagnetic state.   Thus, as the
Zeeman energy is varied compared to the symmetric-antisymmetric
splitting there is a transition from a singlet to a ferromagnetic
state, as shown in Figure 3.

\sec{$\nu=2$ states} Again we now imagine extending our double
well with two electrons to a two dimensional bilayer.  Including
spin, at each position $X$ in space, there are now four possible
state that we can fill, and if we fix our attention to the filling
fraction $\nu=2$ we now have precisely two electrons per position
$X$.  In analogy with the case of two electrons in a double well
(See Fig 3) we might imagine having only two phases, the
ferromagnetic phase at high ratio of Zeeman energy to
symmetric-antisymmetric splitting and the singlet phase at low
ratio of Zeeman energy to symmetric-antisymmetric splitting.  The
wavefunction of the ferromagnetic phase naturally can be written
as
\begin{equation}
    ~~~~~~~|{\bf F}\rangle = \prod_X c^\dagger_{L\downarrow X}c_{R\downarrow X}^\dagger
    |0\rangle
    \label{eq:ferr}
\end{equation}
whereas the wavefunction in the singlet phase is written as
\begin{equation}
\label{eq:Sphase} |{\bf S}\rangle \!= \! \prod_X (u
c^\dagger_{L\downarrow X} \!\!+ v c_{R\downarrow X}^\dagger\!) (u
c^\dagger_{L\uparrow X} \!\!+ v c_{R\uparrow X}^\dagger\!)
    |0\rangle
\end{equation}
where $(u,v)$ is the iso-spinor representing the symmetric
superposition.

Were there not electron-electron interactions, the $\bf S$ and
$\bf F$ phases would be the entire story. However, in the presence
of interactions, near the transition point between these two
phases, it turns out that other phases
exist\cite{Zheng,brey:99,macdonald:99,burkov:02,Anna}, including
so-called ``canted" phases, where the spins in the two layers are
partially aligned --- I.e., the spins are neither fully aligned as
in the $\bf F$ phase, nor fully antialigned as in the $\bf S$
phase.  Spectroscopic experiments have indeed seen very suggestive
indications of phase transitions in $\nu=2$
bilayers\cite{Pellegrini}.

\sec{The Interlayer Coherent $\bf I$-Phase:} Once again, we will
focus our attention on the limit of vanishing tunnelling between
the layers. Reiterating, if $V_z$ is large enough, then both spins
will be in the spin-down state, and we will have the ferromagnetic
phase (Eq. \ref{eq:ferr}).    On the other hand, if energy
difference between the two wells $V_\Delta$ is large enough, then
both of the electrons will go into a single well and we will have
a trivial singlet phase with wavefunction
\begin{equation}
    ~~~~~~~|{\bf S}\rangle = \prod_X c^\dagger_{R\downarrow X}
c^\dagger_{R\uparrow X}
    |0\rangle  \label{eq:t0S}
\end{equation}
where we have assumed here that the right well ($R$) has the lower
of the two potentials.  Note, Eq. \ref{eq:t0S} is just the $t
\rightarrow 0$ limit of the $\bf S$ phase in Eq. \ref{eq:Sphase}
since what we would call the ``symmetric" state (the lower of the
two single particle eigenstates) is just the electron residing in
the lower of the two wells in this limit.

As suggested above, the transition between the $\bf F$ and $\bf S$
phase is not direct.   An interesting phase also
occurs\cite{brey:99} for intermediate values of $V_z /V_\Delta$
which we call the $\bf I$-phase which stands for ``Interlayer
Coherent" phase\footnote{The $\bf I$ phase is the $t \rightarrow
0$ limit of a canted phase\cite{brey:99,macdonald:99}.}. The
wavefunction of the $\bf I$ phase is given by
\begin{equation}
~~~~~~~ |{\bf I}\rangle =  \prod_X (\tilde u
c^\dagger_{L\downarrow X} + \tilde v c_{R\uparrow X}^\dagger)
c^\dagger_{R \downarrow X} |0\rangle
\end{equation}
where the spinor $(\tilde u,\tilde v)$ here allows continuous
interpolation between the $\bf F$ phase ($\tilde u=1, \tilde v =
0)$ and the $\bf S$ phase ($\tilde u=0, \tilde v=1$).

This wavefunction is clearly quite analogous to the $\nu=1$
wavefunction Eq. \ref{eq:111}, and much of the physics is also the
same.   As in that case, here there is a (Goldstone) phase freedom
(the phase of $\tilde v/\tilde u$), and an order parameter
\begin{equation}
~~~~~~~    \psi_X =\langle c^\dagger_{L\downarrow X} c_{R\uparrow
    X}^{\phantom{\dagger}}
    \rangle
\end{equation}
As in the $\nu=1$ case, this order parameter indicates interlayer
coherence that is ``spontaneous" in the  sense that it would not
occur without electron-electron interactions.

It is worth noting, however, that the coherent interlayer
tunnelling current here, \begin{equation}~~~~~~j \sim
 -i ( c^\dagger_{L\downarrow X} c_{R\uparrow
    X}^{\phantom{\dagger}} - c^\dagger_{R\downarrow X}
c_{L\uparrow
    X}^{\phantom{\dagger}}) \end{equation}  analogous to Eq. \ref{eq:intercurr} involves a
spin-flip, and therefore should be extremely suppressed in
experiments. To understand this suppression we need only realize
that, in the absence of spin-orbit terms, which are usually weak,
the total spin angular momentum of the system is precisely
conserved which completely forbids spin-flip tunnelling current.
When one adds back in the weak spin-orbit terms, the spin angular
momentum is ``almost" conserved meaning that spin-flips can only
occur very occasionally --- which strongly limits the interlayer
spin-flip current even if there is such a low-bias Josephson-like
contribution.

\sec{Exciton Language}

Analogous to the $\nu=1$ wavefunction Eq. \ref{eq:111} the $\bf
I$-state can be written in the language of excitons.  We can write
the wavefunction for the $\bf I$-phase in either of two forms. The
first form is built on top of the ferromagnetic phase (Eq.
\ref{eq:ferr}
\begin{equation}
  ~~~~~~~  |{\bf I}\rangle = \prod_X (\tilde u
 + \tilde v c_{R\uparrow X}^\dagger c^{\phantom{\dagger}}_{L\downarrow X} )
 |{\bf F} \rangle
\end{equation}
and the exciton $b^\dagger_X =c_{R\uparrow X}^\dagger
c^{\phantom{\dagger}}_{L\downarrow X}$ now carries spin.  The
other equivalent possibility is to build the state on top of the
singlet phase (Eq. \ref{eq:t0S})
\begin{equation}
~~~~~~~    |{\bf I}\rangle = \prod_X (\tilde u c_{L\downarrow
X}^\dagger c^{\phantom{\dagger}}_{R\uparrow X}
 + \tilde v )
 |{\bf S} \rangle
\end{equation}
where the exicton-boson is then $\tilde b^\dagger_X =
c_{L\downarrow X}^\dagger c^{\phantom{\dagger}}_{R\uparrow X}  =
b_X$ which again is just the conjugate of the above exciton
definition. These two possibilities are, of course, analogous to
the two ways of writing the $\nu=1$ state (Eq. \ref{eq:BCS} and
\ref{eq:BCS2}). In either language, the excitons form a coherent
state, and the superfluid mode of these particles (the Goldstone
mode of the phase degree of freedom of $\tilde v/\tilde u$)
involves counterpropagating currents {\it of opposite spins} in
the opposite layers.  As of yet, there have been no experiments to
try to observe such a superfluid current.

\sec{First Quantized Language}

Finally, it is worth pointing out that the $\bf I$-phase has a
very simple first quantized form analogous to Eq.
\ref{eq:111realspace} above.  We define $z_{\downarrow i}$ with
$i=1, \ldots, N$ to be the positions of the spin-down electrons in
the right layer, $z_{\uparrow i}$  with $i=1, \ldots, N_L$ to be
the positions of the spin-up electrons in the right layer, and
$w_{\downarrow i}$ with $i=1, \ldots, N_R$ to be the positions of
the spin-down electrons in the left layer.  Here again $N = N_R +
N_L$ is the total number of positional states in the lowest Landau
level.   Reasoning analogous to that of section \ref{sec:first}
above now yields the wavefunction
\begin{eqnarray} \nonumber ~~~\Psi_{N_R} &=&   \prod_{i <j
\leq N}
 (z_{\downarrow i} - z_{\downarrow j}) \\ & & \nonumber \prod_{i <j \leq N_R}
 (z_{\uparrow i} - z_{\uparrow j})  \prod_{ i <j \leq N_L}
(w_{\downarrow i} - w_{\downarrow j})  \\ & &   \prod_{i \leq N_R
; j \leq N_L} (z_{\uparrow i} - w_{\downarrow j})
\end{eqnarray}
Here, the first term (the first line) gives a filled Landau level
of spin down electrons in the right layer, and the terms on the
second and third line are analogous to the pieces of the
wavefunction of Eq. \ref{eq:111realspace}.   As required by Fermi
statistics, the wavefunction vanishes whenever two electrons with
the same spin state come to the same position in the same layer
(if any two $z_{\downarrow}$'s come to the same position or if any
two $z_{\uparrow}$'s come to the same position, or if any two
$w_{\downarrow}$'s come to the same position).  Again, the
interesting piece of this wavefunction is the final term which
forces the wavefunction to also vanish if an up-spin electron in
one layer $z_\uparrow$ comes to the same position as a down-spin
electron  in the opposite layer $w_\downarrow$ --- which is just
the up-spin electron binding to the down-spin hole in the opposite
layer.

\section{The Tip of The Iceberg}

The physics discussed above in this paper just scratches the
surface of the interesting physics that can occur in bilayers
--- the possible phase space of states that one can explore in this
system is quite large. In the current paper we have looked at
$\nu=1$ and $\nu=2$ and considered mainly the limit of $t
\rightarrow 0$.  We have considered (albeit briefly) variation of
several parameters including the interlayer bias $V_z$, the Zeeman
energy $V_z$, and the distance between the two layers.  More
generally, one could imagine exploring a much bigger phase space
where we also imagine varying the filling fraction $\nu$, the
tunnelling $t$, the Zeeman energy $V_z$, the interlayer bias
$V_\Delta$, the temperature, the disorder, the in-plane magentic
field\footnote{ Once the tunnelling is nonzero, magnetic field
directed {\it in the plane} of the sample can greatly change the
nature of ground state, since electrons accumulate Aharanov-Bohm
phases when they tunnel between the layers.} as well as a number
of other less obvious experimental parameters (For example, we
could tweak the effective interaction by changing the quantum well
width, or we could change the spin-orbit coupling by using holes
rather than electrons). Some pieces of this larger phase space
have been studied already[1-18], but others still are waiting for
more complete analysis.

It is worth emphasizing, however, that the interest in bilayers
does not rest solely on the fact that there are so many parameters
to play with.  Much more important is the fact that the physics of
these bilayer systems is new and exciting --- blending the effects
of quantum mechanics, statistical physics, and many body
interactions in unique and novel ways.  Over the last decade,
experiment and theory have had a rare and delightful synergy in
pushing knowledge forward. It seems quite clear that this will
remain a productive and active field for years to come.





\begin{thebibliography}{00}


\bibitem{dassarma}
See for example chapters by J.~P.~Eisenstein, S.~Girvin, and
A.~H.~MacDonald, in {\it{Perspectives in Quantum {Hall} Effects}}
eds. S. Das Sarma and A. Pinczuk, Wiley, NY (1997).

\bibitem{TunExp}   I. B. Spielman, J. P.
Eisenstein, L. N. Pfeiffer, and K. W. West Phys. Rev. Lett. {\bf
84}, 5808(2000); Phys. Rev. Lett. {\bf 87}, 036803 (2001).

\bibitem{Exp2}  M. Kellogg, I. B.
Spielman, J. P. Eisenstein, L. N. Pfeiffer, and K. W. West Phys.
Rev. Lett. {\bf 88}, 126804 (2002)

\bibitem{Kellogg2}  M. Kellogg, J.P. Eisenstein, L.N. Pfeiffer, and K.W.
West, Phys. Rev. Lett. {\bf 93}, 036801 (2004); E. Tutuc, M.
Shayegan, and D. A. Huse, Phys. Rev. Lett. {\bf 93}, 036802
(2004).


\bibitem{Pellegrini} V. Pellegrini, A. Pinczuk, B. S. Dennis, A.
S. Plaut, L. N. Pfeiffer, and K. W. West, Science, {\bf 281},
  799, (1998); Phys. Rev. Lett, {\bf 78},
310 (1997).

\bibitem{WenZee}  X.-G. Wen and A. Zee, Phys. Rev. Lett. {\bf 69}, 1811 (1992).

\bibitem{Josephson} Y. N. Joglekar and A. H. MacDonald, Phys. Rev.
Lett. {\bf 87}, 196802 (2001).h

\bibitem{Halperin} B. I. Halperin, Helv. Phys. Acta. {\bf 56}, 75
(1983).


\bibitem{Bonesteel} N. E. Bonesteel, I. A. McDonald, and C. Nayak
Phys. Rev. Lett. {\bf 77},  (1996)


\bibitem{OtherApproaches}  John Schliemann, S.M. Girvin, A.H. MacDonald, Phys. Rev. Lett.
{\bf 86}, 1849 (2001); A. A. Burkov and A. H. MacDonald Phys. Rev.
B 66, 115320 (2002).

\bibitem{Yoshioka} K. Nomura and D. Yoshioka,  Phys. Rev. B {\bf 66}, 153310
(2002);

\bibitem{Us1} S. H. Simon, E. H. Rezayi, and M. V. Milovanovic, Phys. Rev. Lett. {\bf 91}, 046803 (2003)

\bibitem{Gunnar} G. Moller and S. H. Simon, to be published.

\bibitem{Zheng} L. Zheng, R. Radtke, and S. Das Sarma, Phys. Rev. Lett
{\bf 78}, 2453, (1997); S. Das Sarma, S. Sachdev, and L. Zheng,
Phys. Rev. B, {\bf 58}, 4672,  (1998).

\bibitem{brey:99}
L. Brey, E. Demler, and S.Das Sarma,  Phys. Rev. Lett. {\bf 83},
168  (1999).

\bibitem{macdonald:99}
A.~H. MacDonald,
 R. Rajaraman,
 and T. Jungwirth,
 Phys. Rev. B, {\bf 60},
 8817, (1999).


\bibitem{burkov:02}
A.~A. Burkov and A.~H. MacDonald,
 Phys. Rev. B, {\bf 66},
115323, (2002).

\bibitem{Anna}
A.Lopatnikova, S. H. Simon, and E. Demler,  Phys. Rev. B {\bf 70},
in press (2004).

\bibitem{ExcitonCondensate} L. V. Keldysh and Y. V. Kopaev, Fiz.
Tverd. Tela {\bf 6} 2791 [1965 Sov. Phys. Solid State {\bf 6}
2219].  For a modern discussion see P. B. Littlewood et al,  J.
Phys.: Cond. Mat {\bf 16}  S3597 (2004).

\bibitem{Composite} See ``Composite Fermions", ed. O. Heinonen,
World Scientific, 1998.

\bibitem{Read} N. Read, Semi. Cond. Sci. Tech. {\bf 9} 1859
(1994); Surf. Sci. {\bf 361} 7 (1996).







\end{thebibliography}
\end{document}